\documentclass[12pt]{article}
\usepackage{oldgerm}
\usepackage{yfonts}
\usepackage{euler}
\usepackage{graphics}
\usepackage{bm}
\usepackage{graphicx}
\usepackage{amsmath}
\usepackage{amssymb}
\usepackage{amstext}
\usepackage{amscd}
\usepackage{amsfonts}
\usepackage{appendix}
\usepackage{wasysym}
\usepackage{epstopdf}
\usepackage{color}
\DeclareMathAlphabet{\EuFrak}{U}{euf}{m}{n}
\DeclareMathAlphabet{\EuScript}{U}{eus}{m}{n}

\newcommand{\nd}{\noindent}

\title{{\bf The Tachyon Field below the mass barrier}
\thanks{\it{This work was partially supported by Consejo Nacional de
Investigaciones Cient\'{\i}ficas and Comisi\'{o}n de Investigaciones
Cient\'{\i}ficas de la Pcia. de Buenos Aires; Argentina.}}}
\author{{D. G. Barci, C. G. Bollini, M.C.Rocca}, \\
\small{Departamento de F\'{\i}sica,
Universidad Nacional de La Plata,}\\
\small{C.C 67, 1900, La Plata, Argentina}}

\date{December 10, 1992}

\begin{document}

\maketitle

\vspace{-5mm}

\begin{abstract}

We consider a tachyon field whose Fourier components 
correspond to spatial momenta with modulus smaller
than the mass parameter. The plane wave solutions 
have them a time evolution which is a real exponential.
The field is quantized and the solution of the eigenvalue problem
for the Hamiltonian leads to the evaluation of the vacuum
expectation value of products of field operators.
The propagator turns out to be half-advanced and half-retarded.
This completes the proof [4] that the total propagator
is the Wheeler Green function [4,7].
\nd

\end{abstract}

\newpage

\section{Introduction}

For a normal particle (bradyon), the Klein-Gordon equation
\begin{equation}
\label{eq1}
(\Box+m^2)\varphi=0
\end{equation}
leads to mass-energy relation of the form
\begin{equation}
\label{eq2}
E=(\vec{k}^2+m^2)^{\frac {1} {2}}
\end{equation}
which implies that $E$ is real for any real value of $\vec{k}$.

Instead, for the tachyon field we have
\begin{equation}
\label{eq3}
(\Box-\mu^2)\phi=0
\end{equation}
so that the elementary solution is
\begin{equation}
\label{eq4}
\phi=e^{i(wt-\vec{k}\cdot\vec{x})}\;\;\;\;\;\rm{with}\;\;\;\;\;
w=(\vec{k}^2-\mu^2)^{\frac {1} {2}}
\end{equation}
and $w$ is real only when $\vec{k}^2\geq\mu^2$.
For $\vec{k}$ below the mass barrier ($\vec{k}^2<\mu^2$),
$w$ is pure imaginary and the exponential in [4]
blows up either for $t\rightarrow -\infty$ or for 
$t\rightarrow+\infty$.In general this region of $\vec{k}$
is discarded (See Ref. [1] and [2]) and only values such that
$\vec{k}^2\geq\mu^2$ is accepted.

However the situation here is similar to the case of the ordinary
quantum-mechanical description of the movement of a free particle
that find in its way a potential barrier higher than its kinetic energy.
The real solutions inside the potential cannot be ignored.
In our case the ''mass'' squared $\mu^2$ plays the role of the
potential barrier.

It is nowadays that tachyons cannot appear asymptotically as
free particles without impairing unitarity [3], but this does not
mean that they have to be barred from consideration.
They should be forbidden to propagate freely but they can be allowed
to exist as transient virtual modes.

This point of view was adopted in a previous paper for
$\vec{k}^2\geq\mu^2$ [4]. Now we are going to examine some
consequences of considering (3) and (4) for
$\vec{k}^2<\mu^2$.

It is perhaps convenient at this point to warm about a possible
misinterpretation of (4). It should be clear that an imaginary 
value of $w$ means that this quantity cannot be the energy
of any state of the field. As matter of fact we will see that,
even if the exponential in (4) is a solution of the field equation (3),
there is no actual eigenstate with particle-like properties
which can be considered similar to those of bradyons.

In this paper only values of $\vec{k}$ such that 
$\vec{k}^2<\mu^2$ will be considered.

\section{Quantization}

Any real solution of (3) can be written as
\begin{equation}
\label{eq5}
\phi=\phi_1+\phi_2
\end{equation}
\begin{equation}
\label{eq6}
\phi_1=\frac {1} {(2\pi)^{\frac {N-1} {2}}}
\int\frac {dk} {\sqrt{w}}\left(b_k^1e^{-wt}+c_k^1e^{wt}\right)
\cos\vec{k}\cdot\vec{r}
\end{equation}
\begin{equation}
\label{eq7}
\phi_1=\frac {1} {(2\pi)^{\frac {N-1} {2}}}
\int\frac {dk} {\sqrt{w}}\left(b_k^2e^{-wt}+c_k^2e^{wt}\right)
\sin\vec{k}\cdot\vec{r}
\end{equation}
where from now on we define
\begin{equation}
\label{eq8}
w=|(\vec{k}^2-\mu^2)^{\frac {1} {2}}|=
(\mu^2-\vec{k}^2)^{\frac {1} {2}}
\end{equation}
and all integrations are to be carried out over the region
$\vec{k}^2<\mu^2$.

The Lagrangian for the field is:
\begin{equation}
\label{eq9}
{\cal L}=\frac {1} {2}\partial_\mu\phi\partial^\mu\phi
+\frac {1} {2} \mu^2\phi^2
\end{equation}
The total Hamiltonian can be written
\begin{equation}
\label{eq10}
{\cal H}={\cal H}_1+{\cal H}_2
\end{equation}
\begin{equation}
\label{eq11}
{\cal H}_j=\int dk\frac {w} {2}(b_k^jc_k^j+
c_k^jb_k^j)
\end{equation}
It is now easy to see that canonical quantization of the field
leads to the commutation relations
\begin{equation}
\label{eq12}
[b_k^l,c_{k^{'}}^m]=i\delta^{lm}\delta(\vec{k}-\vec{k}^{'})
\end{equation}
The eigenvalue problem for ${\cal H}_j$ has been discussed
in Ref. [5]. For the sake of completeness we shall give some
results that will be needed afterwards.

For each degree of freedom (each $\vec{k}$), the Hamiltonian has the form
\begin{equation}
\label{eq13}
h=\frac {1} {2}(qp+pq)=qp-\frac {i} {2}\;\;\;\;\;[q,p]=i
\end{equation}
The eigenvalue equation is
\begin{equation}
\label{eq14}
h\psi=\left(qp-\frac {i} {2}\right)\psi=E\psi
\end{equation}
or, in the usual coordinate representation
\begin{equation}
\label{eq15}
-iq\frac {d} {dq}\psi=\left(E+\frac {i} {2}\right)\psi
\end{equation}
whose solution we write as
\begin{equation}
\label{eq16}
\psi^E=\frac {1} {\sqrt{2\pi}}
q_+^{iE-\frac {1} {2}}\;\;\;\;\;-\infty<E<+\infty
\end{equation}
[$q_{+}^{\alpha}=q^{\alpha}\Theta(q)$
and similar solutions with
$q_{-}^{\alpha}=|q|^{\alpha}\Theta(-q)$]
The spectrum is continuous and runs the real energy axis
from $-\infty$ to $+\infty$. The normalization is such that:
\[(\psi^E,\psi^{E^{'}})=\frac {1} {2\pi}\int\limits_0^{\infty}
dq q^{-iE-\frac {1} {2}}q^{iE^{'}-\frac {1} {2}}=
\frac {1} {2\pi}\int\limits_0^{\infty}\frac {dq} {q}
q^{i(E^{'}-E)}=\]
\begin{equation}
\label{eq17}
\frac {1} {2\pi}\int\limits_{-\infty}^{\infty} dy\;
e^{i(E^{'}-E)y}=\delta(E-E^{'})
\end{equation}
($y=\ln q$)
The solutions (16) are then similar to the plane wave
solutions for free particles. They have been used in quantum
optics to describe the time solutions ''squeezed states''
and also to evaluate the Shannon entropy for the degenerate
parametric amplifier whose Hamiltonian can be cast
into the form (11) (See Ref. [13]).

The orthogonality relations are contained (for $E\neq E^{'}$)
in the general expression (see Ref. [6])
\begin{equation}
\label{eq18}
\int\limits_0^{\infty} dq\;q^{\alpha}=0\;\;\;\;\;
(\alpha\neq -1)
\end{equation}
These formulae show that the mean values of $q^{\mu}$
(and $p^\mu$) are 0 for $\mu\neq 0$
\begin{equation}
\label{eq19}
(\psi^E,q^\mu\psi^{E^{'}})=\frac {1} {2\pi}\int\limits_0^{\infty}
dq q^{-iE-\frac {1} {2}}q^\mu q^{iE-\frac {1} {2}}=
\frac {1} {2\pi}\int\limits_0^{\infty}
dq \;q^{\mu-1}=0\;\;\,(\mu\neq 0) 
\end{equation}
On the other hand we have:
\[(\psi^E,pq\psi^{E^{'}})=\frac {1} {2\pi}\int\limits_0^{\infty}
dq q^{-iE-\frac {1} {2}}(-i)\frac {d} {dq}qq^{iE^{'}-\frac {1} {2}}=\]
\begin{equation}
\label{eq20}
\left(E^{'}-\frac {i} {2}\right)\delta(E-E^{'})=
\left(E-\frac {i} {2}\right)
(\psi^E,\psi^{E^{'}})
\end{equation}
and of course
\begin{equation}
\label{eq21}
(\psi^E,qp\psi^{E^{'}})=
\left(E+\frac {i} {2}\right)
(\psi^E,\psi^{E^{'}})
\end{equation}
We shall now write the zero energy mean values using
the notation $\psi^{E=0}\equiv |0>$. Then (19)-(21) reads
\begin{equation}
\label{eq22}
<0|q\mu|0>=0=<0|p\mu|0>\;\;\;\;\;(\mu\neq 0)
\end{equation}
\begin{equation}
\label{eq23}
<0|pq\mu|0>=-\frac {i} {2}<0|0>=-<0|qp\mu|0>
\end{equation}

\section{Vacuum Expectation Values of Products of Field Operators}

Taking into account (6), (12), (22) y (23), we can evaluate the mean value
for products of the $\phi_1$ field at different points (we shall take
$<0|0>=1$)
\[<0|\phi_1(x)\phi_1(y)|0>=\]
\[\frac {1} {(2\pi)^{N-1}}
\int\frac{dk} {\sqrt{w}}\int\frac{dk^{'}} {\sqrt{w^{'}}}
<0|b_k^1c_k^1|0>\cos\vec{k}\cdot\vec{x}
\cos\vec{k}\cdot\vec{y}
e^{-(wx_0-w^{'}y_0)}+\]
\[<0|c_k^1b_k^1|0>\cos\vec{k}\cdot\vec{x}
\cos\vec{k}\cdot\vec{y}
e^{(w^{'}x_0-wy_0)}=\]
\begin{equation}
\label{eq24}
<0|\phi_1(x)\phi_1(y)|0>=
\frac {i} {(2\pi)^{N-1}}
\int\frac{dk} {2w}
\cos\vec{k}\cdot\vec{x}
\cos\vec{k}\cdot\vec{y}
(e^{-w(x_0-y_0)}-
e^{w(x_0-y_0)})
\end{equation}
Analogously
\begin{equation}
\label{eq25}
<0|\phi_2(x)\phi_2(y)|0>=
\frac {i} {(2\pi)^{N-1}}
\int\frac{dk} {2w}
\sin\vec{k}\cdot\vec{x}
\sin\vec{k}\cdot\vec{y}
(e^{-w(x_0-y_0)}-
e^{w(x_0-y_0)})
\end{equation}
Adding (23) to (24) we get
\begin{equation}
\label{eq26}
<0|\phi(x)\phi(y)|0>=
\frac {i} {(2\pi)^{N-1}}
\int\frac{dk} {2w}
e^{-i\vec{k}\cdot(\vec{x}-\vec{y})}
(e^{-w(x_0-y_0)}-
e^{w(x_0-y_0)})
\end{equation}
from which we obtain
\[<0|T\phi(x)\phi(y)|0>=\]
\[\frac {i} {(2\pi)^{N-1}}
\int\frac{dk} {2w}
e^{-i\vec{k}\cdot(\vec{x}-\vec{y})}
(e^{-w(x_0-y_0)}-
e^{w(x_0-y_0)})Sg(x_0-y_0)\]
\begin{equation}
\label{eq27}
\frac {i} {(2\pi)^{N-1}}
\int\frac{dk} {2w}
e^{-i\vec{k}\cdot(\vec{x}-\vec{y})}
\sinh|x_0-y_0|=
\frac {1} {2\pi}W(x-y)
\end{equation}
The Green function (27) is the Wheeler propagator defined
in Ref. [7]. Further if we note that (Ref. [8], 3.354, p. 312)
\[\int\limits_{-\infty}^{\infty}
\frac {e^{-ik_0t}} {k_0^2+w^2}=
\frac {\pi} {w}e^{-wt}\]
then we see that $W(x)$ is the ''Fourier transform'' of
$(k_0^2+w^2)^{-1}=(k_0^2-\vec{k}^2+\mu^2)^{-1}$
where the $k_0$ integration path can be considered to be
the superposition of two branches. In one of them both
poles are left below the path integration and in the other
branch are left above the path. In other words, $W(x)$
is half-retarded plus half-advanced (see also Ref. [9] and
[10]) 
\begin{equation}
\label{eq28}
W(x)=\frac {1} {2}\Delta_R(x)+\frac {1} {2}\Delta_A(x)
\end{equation}

\section{Lorentz Invariance}

The generators of the Poincar\'e group can be constructed
of the energy-momentum tensor:
\begin{equation}
\label{eq29}
{\cal T}_\mu^\nu=-\partial_\mu\partial_\nu\phi+
\frac {1} {2}(\partial_\rho\partial_\rho\phi
-\mu^2\phi^2)\delta_\mu^\nu
\end{equation}
With (5)-(7) we find that
\[{\cal P}_0=\frac {i} {2}\int dk\; k_0
[\{b_k^1,c_k^1\}+\{b_k^2,c_k^2\}]\;\;\;\;\;k_0=iw\]
\begin{equation}
\label{eq30}
{\cal P}_i=\frac {1} {2}\int dk\; k_i
[\{b_k^2,c_k^1\}-\{b_k^1,c_k^2\}]\;\;\;\;\;1\leq i\leq 3
\end{equation}
From the angular momentum tensor
\begin{equation}
\label{eq31}
{\cal M}_{\mu\nu}^\rho=x_\mu{\cal T}_\nu^\rho-
x_\nu{\cal T}_\mu^\rho
\end{equation}
we obtain, in particular, the Lorentz boost
\begin{equation}
\label{eq32}
{\cal M}_{0i}=\frac {1} {2}\int dk\; w
[\{\partial_ib_k^2,c_k^1\}-\{\partial_ib_k^1,c_k^2\}]\;\;\;\;\;
\partial_i\equiv\frac {\partial} {\partial k_i}
\end{equation}
It is a matter of tedious algebra to verify that the operators
${\cal P}_\mu$ and ${\cal M}_{\mu\nu}$ obey the commutation
relations of teh Poincar\'e algebra.

To verify that $\phi$ transforms as a scalar under the 
transformations of the Lorentz group generated by 
${\cal M}_{\mu\nu}$, it is necessary to take into account
that the Fourier transform of $\phi$ is an analytic generalized
function (analytic distribution; see Ref. [11]), so that the path
of integration [in (6) and (7)] can be deformed in the complex
$\vec{k}$ plane. But this proof is actually not needed, as the
field of the tachyon disappears from all matrix elements
(in contradistinction to the case of bradyons) For this reason
the tachyonic field should be considered more an auxiliary
concept than a real entity. The only trace of the field $\phi$
leaves is the Wheeler propagator, which, as explained in Sec. 3
is the Fourier transform of
$(k_0^2-\vec{k}^2+\mu^2)^{-1}=(k^2+\mu^2)^{-1}$

The complete propagator is Lorentz invariant, as shown
by K. Kamoi and S. Kamefuchi in Ref. [12] for the
retarded and advanced Green function (cf. eq. (28)).

\newpage

\section{Discussion}

When tachyons are forbidden to propagate asymptotically
as free particles, all values of $\vec{k}$ are acceptable in
principle and $\vec{k}$ space divides naturally into two
regions: (a) the sphere $\vec{k}^2<\mu^2$ and (b) the
complement $\vec{k}^2\geq\mu^2$. The case (b) was considered
in Ref. [4]. There it was shown that the coefficients of 
elementary solutions are increasing and decreasing operators.
None of them annihilates the vacuum, which is an eigenstate
of the Hamiltonian with zero energy and zero momentum
When $\vec{k}^2>\mu^2$ and $\vec{k}^2\rightarrow\mu^2$
the energy momentum spectrum is discrete but becomes
more and more dense. For $\vec{k}^2<\mu^2$ we find a 
continuum. For (a) or (b), the propagator is found to be
half-advanced plus half-retarded, the only difference being
that in (b) both poles in the $k_0$ plane are real while in (a)
they are pure imaginary.

The tachyon field is related to complex mass fields through
the Wheeler propagator, which appears in a natural way
(see also Ref. [7]).

Furthermore, this is practically, the only Green function
which is compatible with the suppression of the asymptotic
free modes of the field.

\end{document}